\documentclass[preprint]{elsarticle}
\usepackage[T1]{fontenc}
\usepackage[pdftex]{color}
\usepackage{amsmath,amssymb,amsthm,eucal,upref}

\date{}

\theoremstyle{plain}
\newtheorem{theorem}{Theorem}[section]

\theoremstyle{definition}

\theoremstyle{definition}

\numberwithin{equation}{section}

\newcommand{\RR}{\mathrm{I\kern-0.20emR}}
\newcommand{\E}{\mathrm{e}\kern0.2pt}%

\newcommand{\be}{\begin{equation}}
\newcommand{\ee}{\end{equation}}

\newcommand{\HH}{H}

\begin{document}

\title{{\bf Small-amplitude steady water waves with critical layers: non-symmetric waves}}

\author[vk]{V.~Kozlov\corref{cor1}}
\ead{vladimir.kozlov@liu.se}
\author[el]{E.~Lokharu\corref{cor2}}
\ead{evgeniy.lokharu@math.lu.se}

\address[vk]{Department of Mathematics, Link\"oping University, S--581 83 Link\"oping}
\address[el]{Department of Mathematics, Lund University, S--223 62 Lund}

\begin{abstract}
	
	The problem for two-dimensional steady water waves with vorticity is considered. Using methods of spatial dynamics, we reduce the problem to a finite dimensional Hamiltonian system. The reduced system describes all small-amplitude solutions of the problem and, as an application, we give a proof of the existence of non-symmetric steady water waves whenever the number of roots of the dispersion equation is greater than one.
	
\end{abstract}

\maketitle

\section{Introduction}

In the present paper we consider the problem describing two-dimensional steady water waves with vorticity under the action of gravity. The fluid is assumed to be incompressible and we neglect the effects of the surface tension. We will be interested in the analysis of small-amplitude waves over streams with counter-currents. A similar setup appeared in a several recent papers including \cite{Wahlen09} and \cite{Kozlov2014} on the existence of Stokes waves with critical layers, \cite{EhrnstromEscherWahlen11} and \cite{Aasen2017} concerning the existence of bimodal waves, and \cite{EhrnstromWahlen15}, \cite{Kozlov2017} constructing trimodal and N-modal waves respectively. In all mentioned papers the authors prove existence using a local bifurcation argument of Crandall and Rabinowitz (see \cite{CrandallRabinowitz71}) and its generalizations to the case of multi-dimensional kernels. A distinct feature of the method is the usage of certain bifurcation parameters of the problem such as mass flux or total head to prove existence of families of small solutions for which these parameters are varying along the bifurcation curve. In general such approach does not provide enough information on the structure of the set of small-amplitude waves for fixed values of problems parameters. In this paper we use a different approach which is known as spatial dynamics, which allows us to reformulate the water wave problem as a finite-dimensional Hamiltonian system, for which the Hamiltonian given by the flow force invariant is negative definite. The latter implies  existence of a large class of small waves and we will prove that most of them are non-symmetric.

The first Hamiltonian formulation of steady Euler equations is due to Zakharov \cite{Zakharov68} in 1968. Since then many studies had been done on variational formulations and Hamiltonian structures of the problem in the irrotational case of zero vorticity both with and without surface tension. Some of them are \cite{BENJAMIN1984},\cite{Radder1992}, \cite{GrovesStylianou14} treating different Hamiltonian formulations. However that Hamiltonian formulations could not be used directly for proving existence of solutions since they were infinite-dimensional. That was true before the paper \cite{Kirchgaessner88} by Kirchg\"assner published in 1988, where a new approach was introduced. The idea of Kirchg\"assner was to use some form of centre-manifold reduction to reformulate the problem as a finite system of ordinary differential equations. Then a proper analysis of the reduced equations has lead to the first existence proofs for gravity capillary solitary waves (\cite{AmickKirchgaessner89} and \cite{IoossKirchgaessner90}; see also \cite{BuffoniGrovesToland96}). The method of Kirchg\"assner is now known as spatial dynamics. A naive idea of the method is to formulate an infinite- dimensional problem as a dynamical system, where the role of time is played by a spatial unbounded coordinate. For two-dimensional steady water waves it would be $x$-coordinate measured in the direction of propagation. Then, under certain assumptions, the problem can be reduced to a finite-dimensional system of ordinary differential equations inheriting many useful properties such as Hamiltonian structure and reversibility. The first results related to the method (see \cite{Kirchgaessner82}, \cite{Mielke1986}, \cite{Mielke88}) covered the case of some elliptic problems in cylinders. A general case of the reduction is treated in \cite{Mielke1991}, where it is proved that the centre-manifold reduction method of Kirchg\"assner preserves the Hamiltonian structure. A general theorem providing the center-manifold reduction which is the major tool of spatial dynamics and some of its generalizations are due to Mielke \cite{Mielke1991}.

Recently the method of spatial dynamics was successfully used in \cite{GrovesWahlen07} and \cite{GrovesWahlen08} for proving the existence of two-dimensional solitary type waves with and without surface tension for an arbitrary vorticity distribution. For applications of the method to three-dimensional waves we refer to \cite{GrovesMielke01}, \cite{GrovesHaragus03}, \cite{Groves2018}, \cite{BuffoniGrovesSunWahlen13}, \cite{GrovesSunWahlen16a}. \\

In this paper we use spatial dynamics technique for two-dimensional waves of small amplitude over flows with counter currents. In contrast to the previous results we do not restrict ourselves to the case of near-critical values of the problems parameters and we also allow interior stagnation. In this case it is possible for the centre-manifold to have an arbitrary dimension. We show that whenever the dispersion equation has a root (or, equivalently, the centre-manifold is not trivial), then the problem can be reduced to a finite-dimensional Hamiltonian system with the Hamiltonian given by the flow force invariant. The first quadratic and cubic terms of the Hamiltonian are given explicitly. Because the reduced Hamiltonian is negative-definite, the problem possesses a large class of small-amplitude solutions. We prove that if there two or more roots of the dispersion equation, then the most of the small solutions are non-symmetric. The existence of non-symmetric waves was predicted numerically (see \cite{Wang2014}, \cite{Aider2004}) in the presence of surface tension on the deep water.

The paper is organized as follows. In Section 2 we formulate the problem and in Section 3 we state our main result Theorem \ref{T1}. Our argument is based on the application of centre-manifold reduction theorem due to Mielke \cite{Mielke1991}. One of the main difficulties is the nonlinear boundary condition (Bernoulli equation). In Subsection 3.4 we eliminate this difficulty by a proper change of variables, which is technical but standard (see \cite{GrovesWahlen08}). Finally, Section 4 is devoted to the existence of non-symmetric waves.

\section{Statement of the Problem}

Let an open channel of uniform rectangular cross-section be bounded below by a
horizontal rigid bottom and let water occupying the channel be bounded above by a
free surface not touching the bottom. In appropriate Cartesian coordinates $(X,Y)$,
the bottom coincides with the $X$-axis and gravity acts in the negative
$Y$-direction. The steady water motion is supposed to be two-dimensional and rotational; the
surface tension is neglected on the free surface of the water, where the pressure is
constant. These assumptions and the fact that water is incompressible allow us to
seek the velocity field in the form $(\psi_Y, -\psi_X)$, where $\psi (X,Y)$ is
referred to as the {\it stream function}. The vorticity distribution $\omega$ is
supposed to be a prescribed smooth function depending only on the values of $\psi$. We choose the frame of reference so that the velocity field is time-independent as
well as the unknown free-surface profile. The latter is assumed to be the graph of
$Y = \eta (X)$, $X \in \Bbb R$, where $\eta$ is a positive continuous function, and
so the longitudinal section of the water domain is ${\cal D} = \{ X \in \Bbb R , \ 0 < Y <
\eta (X) \}$. Under the given assumptions we obtain the following non-dimensional free-boundary problem for $\psi$ and $\eta$:
\begin{eqnarray}
&& \psi_{XX} + \psi_{YY} + \omega (\psi) = 0, \quad (X,Y) \in {\cal D} ; \label{eq:lapp} \\
&& \psi (X,0) = 0, \quad X \in \Bbb R ; \label{eq:bcp} \\ && \psi (X,\eta (X)) = m,
\quad X \in \Bbb R ; \label{eq:kcp} \\ && |\nabla \psi (X,\eta (X))|^2 + 2 \eta (X) = 2Q, \quad X \in \Bbb R . \label{eq:bep}
\end{eqnarray}
In condition \eqref{eq:bep} (Bernoulli's equation), $Q$ is a constant referred to as Bernoulli's constant and $m$ is the mass flux. In what follows we will assume that $\omega \in C^{\nu+2,\gamma}(\RR)$ for some integer $\nu \geq 1$ and $\gamma \in (0,1)$.

In what follows we will assume that $m = m(\lambda),Q = Q(\lambda)$ and $\omega = \omega(\lambda)$ depend on a parameter $\lambda \in \RR^m$ that is taken from an open neighborhood $\Lambda$ of some $\lambda_\star \in \Lambda \subset \RR^m$. The latter defines a family of problems $\textrm{P}_{\psi,\eta}^{\lambda}$ of the form \eqref{eq:lapp}-\eqref{eq:bep}. We will describe below this dependence more precisely. 

By a {\it stream} (shear-flow) solution we mean a pair $(\psi,\eta)=(u (Y), \, d)$, where $u \in C^{ \nu+4,\gamma}([0,d])$ and $d=\mbox{const}$, solving  problem \eqref{eq:lapp}--\eqref{eq:kcp} (for some constant $m \in \RR$) which reduces to the following one:
\[
u'' + \omega (u) = 0 \ \ \mbox{on} \ (0, d) , \ \ \ u (0) = 0, \ \ \ u (d) \neq 0, \label{eq:ss1}
\]
while the corresponding Bernoulli constant and the mass flux are
\begin{equation} \label{Qm}
Q = [u'^2(d) + 2d]/2, \ \ m = u(d).
\end{equation}
A detailed study of these solutions including those that describe flows with counter-currents is given in \cite{Kozlov2011}. We will assume that we are given a family of stream solutions $(u(Y;\lambda),d(\lambda))$ solving the corresponding problem $\textrm{P}_{\psi,\eta}^{\lambda}$, while the functions $\omega$, $u$ and $d$ depend on $\lambda \in \Lambda$ and satisfy
\[
\begin{split}
 & \omega \in C^{\nu+2,\gamma}(\RR \times \Lambda), \ \ d \in C^{\nu+4,\gamma}(\Lambda), \ \ 
  u(\cdot/d,\lambda) \in C^{\nu+4,\gamma}([0,1] \times \Lambda).
\end{split}
\]
Then one can see from the definition \eqref{Qm} that
\[
m \in C^{\nu+4,\gamma}(\Lambda), \ \ \ Q \in C^{\nu+3,\gamma}(\Lambda).
\]
In addition to that, we assume
\[
k(\lambda) := u_{Y}(d;\lambda) \neq 0, \ \ \lambda \in \Lambda.
\]
Finally, when $\lambda = \lambda_\star$, we put
\[
\begin{split}
 & \omega_\star (p)=\omega(p;\lambda_\star), \ \ \ u_\star(Y) = u(Y; \lambda_\star), \ \ \ Q_\star = Q(\lambda_\star), \\
 & d_\star = d(\lambda_\star), \ \ \ m_\star = m(\lambda_\star), \ \ \ k_\star = k(\lambda_\star). 
\end{split} 
\]

For the rest of the paper we will deal with the family of problems $\textrm{P}_{\psi,\eta}^{\lambda}$ supplemented by a family of stream solutions defined above.

\subsection{Scaling}

In what follows it will be convenient to have the depth $d = d(\lambda)$ of the stream solutions to be fixed, that is independent of $\lambda$. This can be achieved by performing the following scaling
\[
\begin{split}
& \tilde{Y} = \frac{d_\star}{d}Y, \ \ \tilde{X} = \frac{d_\star}{d}X, \ \ \tilde{\psi}(\tilde{X},\tilde{Y}) = \left[\frac{d_\star}{d}\right]^{3/2} \psi(X,Y), \\
& \tilde{\eta}(\tilde{X}) = \frac{d_\star}{d} \eta(X), \ \ \tilde{Q}(\lambda) = \frac{d_\star}{d}Q(\lambda), \ \ \tilde{m}(\lambda) = \left[\frac{d_\star}{d}\right]^{3/2} m(\lambda), \\
& \tilde{\omega}(p; \lambda) = \left[\frac{d_\star}{d}\right]^{3/2} \omega\left(\left[\frac{d_\star}{d}\right]^{-3/2} p, \lambda\right).
\end{split}
\]
Thus, every problem $\textrm{P}_{\psi,\eta}^{\lambda}$ transforms into $\tilde{\textrm{P}}_{\tilde{\psi},\tilde{\eta}}^{\lambda}$ corresponding to $(\tilde{\omega}(p;\lambda), \tilde{m}(\lambda), \tilde{Q}(\lambda))$. Furthermore, the family of stream solutions $(u(Y;\lambda),d(\lambda))$ translates into the family $(\tilde{u}(\tilde{Y};\lambda), d_\star)$ solving the problem $\tilde{\textrm{P}}_{\tilde{\psi},\tilde{\eta}}^{\lambda}$. The depth for these new family of stream solutions equals to $d_\star$ and is independent of $\lambda$. Hence, without loss of generality, we will assume that $d = d_{\star}$ is constant.


\subsection{Dispersion equation}

Dispersion relation plays a central role in the theory of small-amplitude steady waves. To see that, we need to derive a linear approximation of the problem \eqref{eq:lapp}-\eqref{eq:bep}. For this purpose it is convenient to rectify the domain $\cal D$ by scaling the vertical variable to
\[
z = Y \frac{d}{\eta (x)},
\]
while the horizontal coordinate remains unchanged: $x = X$. Thus, the domain $\cal D$ transforms into the strip $S = \RR \times (0,d)$. Next, we introduce a new unknown function $\hat{\Phi}  (x, z)$ on
$S$ by
\[
\hat{\Phi} (x, z) = \psi \left( x, \frac{z}{d} \, \eta (x) \right). \label{zeta/phi}
\]
A direct calculation shows that problem \eqref{eq:lapp}-\eqref{eq:kcp} reads in new variables as
\begin{align}
& \left[ \hat{\Phi}_x - \frac{z \eta_x}{\eta} \hat{\Phi}_z \right]_{x} - \frac{z \eta_x}{\eta} \left[ \hat{\Phi}_x - \frac{z \eta_x}{\eta} \hat{\Phi}_z \right]_z + \left( \frac{d}{\eta} \right)^2 \hat{\Phi}_{zz} + \omega(\hat{\Phi} ) = 0; \label{lapp1} \\
& \hat{\Phi}(x,0) = 0, \ \ \ \hat{\Phi}(x,d) = m, \ \ x \in \RR; \label{boc1}
\end{align}
while the Bernoulli equation \eqref{eq:bep} becomes
\begin{equation} \label{bep1}
\hat{\Phi}_z^2 = \frac{\eta^2}{d^2} \left( \frac{2Q - 2 \eta}{1 + \eta_x^2} \right).
\end{equation}

Now we formally linearize equations \eqref{lapp1}-\eqref{bep1} near a stream solution $(u(z;\lambda),d)$.  Thus, we put
\[
\hat{\Phi}(x,z) = u(z;\lambda) + \epsilon \hat{\Phi}^{(1)}(x,z) + O(\epsilon^2), \ \ \eta(x) = d + \epsilon \eta^{(1)}(x)+  O(\epsilon^2).
\]
Using this ansatz in \eqref{lapp1}-\eqref{bep1}, we find after taking the limit $\epsilon \to 0$ the following equations for $(\hat{\Phi}^{(1)},\eta^{(1)})$:
\begin{equation} \label{K1a}
\begin{split}
& \left[ \hat{\Phi}^{(1)}_x - \frac{z \eta^{(1)}_x u_z}{d} \right]_x + \hat{\Phi}^{(1)}_{zz} - \frac{2 \eta^{(1)} u_{zz}}{d} + \omega'(u) \hat{\Phi}^{(1)} = 0, \\
& \hat{\Phi}^{(1)}(x,0) = \hat{\Phi}^{(1)}(x,d) = 0,\\
& \hat{\Phi}^{(1)}_z\vert_{z=d} - \left( \frac{u'(d)}{d} - \frac{1}{u'(d)}  \right) \eta^{(1)} = 0.
\end{split}
\end{equation}
We can simplify equations by letting
\[
\Psi^{(1)} = \hat{\Phi}^{(1)} - \frac{z \zeta^{(1)} u_z}{d}.
\]
The formula above implies $\eta^{(1)} = -\Psi^{(1)}\vert_{z=d} / u'(d)$ and then the problem (\ref{K1a}) transforms into
\begin{align}
& \Psi^{(1)}_{xx}+\Psi^{(1)}_{zz} + \omega'(u) \Psi^{(1)} = 0 \nonumber \\
& \Psi^{(1)}\vert_{z=0} = 0 \label{lin} \\
& \Psi^{(1)}_z\vert_{z=d} - \kappa \Psi^{(1)}\vert_{x=d} = 0, \nonumber
\end{align}
where
\begin{equation} \label{kappadef}
\kappa(\lambda) = 1/[u'(d; \lambda)]^2 - \omega(1; \lambda)/u'(d; \lambda).
\end{equation}
To find bounded solutions of the system \eqref{lin} we can use separation of variables, which leads to the following Sturm-Liouville problem:
\begin{equation} \label{SLdisp}
 - \varphi_{zz}  - \omega' (u)  \varphi = \mu \varphi \ \mbox{on} \ (0, d), \quad
\varphi (0) = 0 , \ \  \, \varphi_z (d) = \kappa \varphi (d).
\end{equation}

The spectrum of the latter eigenvalue problem, referred to as the dispersion equation, consists of a countable set of simple real eigenvalues $\{ \mu_j \}_{j=1}^{\infty}$ ordered so that
$$
\mu_1 < \mu_2<\mu_3 <\cdots.
$$
Furthermore, only a finite number of eigenvalues may be negative.  The normalized eigenfunction corresponding to an eigenvalue $\mu_j$ will be denoted by $\varphi_j$. Thus, the set of all eigenfunctions $\{\varphi_j\}_{j=1}^\infty$ forms an orthonormal basis in $L^2(0,d)$. Note that both eigenvalues $\mu_j = \mu_j(\lambda)$ and eigenfunctions $\phi_j = \phi_j(\cdot;\lambda)$ depend on $\lambda$ and are of order $C^{1+\nu,\gamma}$ in $\Lambda$ and $C^{3+\nu,\gamma}$ in $[0,d] \times \Lambda$ respectively.

Solving the linear problem \eqref{lin}, we find that the space of all bounded solutions is finite-dimensional and is spanned by the functions
\[
\cos(\sqrt{|\mu_j|} x) \varphi_j(z), \ \ \ \sin(\sqrt{|\mu_j|} x) \varphi_j(z),
\]
where $\mu_j \leq 0$ are all non-positive eigenvalues of \eqref{SLdisp}. \\

In addition to the notations of the previous section, we define
\[
\kappa_\star=\frac{1}{k_\star^2}-\frac{\omega_\star(1)}{k_\star}
\]
and
$$
\mu_j^*=\mu_j(\lambda_\star),\;\;\varphi_j^\star(z)=\varphi_j(z,\lambda^\star) \ j\geq 1.
$$
In what follows we will assume that
\begin{equation}\label{KK12a}
\mu_1^*<...<\mu_N^* \leq 0, \ \ \ \mu_{N+1}^* > 0
\end{equation}
for some $N \geq 1$ which will be fixed throughout the paper. We will assume also that $\mu_{N+1}(\lambda)>0$ for $\lambda\in\Lambda$.

\section{Reduction to a finite dimensional system}

\subsection{First order system}

Let us write \eqref{lapp1} as a first order system. For this purpose, we introduce a new variable
\[
\hat{\Psi}(x,z) = \frac{\eta(x)}{d} \left[ \hat{\Phi}_x(x,z) - \frac{z\eta_x(x)}{\eta(x)} \hat{\Phi}_z(x,z) \right].
\]
Thus, writing \eqref{lapp1} in terms of $\hat{\Psi}$ and $\hat{\Phi}$ and using the definition of $\hat{\Psi}$ to express $\hat{\Phi}_x$, we obtain
\begin{align}
& \hat{\Phi}_x = \frac{d}{\eta}\hat{\Psi} + \frac{z}{\eta} \eta_x \hat{\Phi}_z,  \label{tmp1}\\
& \hat{\Psi}_x =  \frac {1}{\eta}\eta_x (z\hat{\Psi})_z - \frac{d}{\eta} \hat{\Phi}_{zz} - \frac{\eta}{d} \omega(\hat{\Phi}) \label{tmp2}.
\end{align}
Moreover, the function $\hat{\Phi}$ satisfies \eqref{boc1} and \eqref{bep1}. In what follows we will assume that $\hat{\Phi}$ and $\hat{\Psi}$ are continuous functions of $x$ with values in the manifolds $\HH^2_{b,m}$ and $\HH^1_b$ respectively, where
\[
\begin{split}
& \HH^n_{b} = \{ \phi \in H^n(0,d): \phi(0)=0 \}, \\
& \HH^n_{b,\theta} = \{ \phi \in H^n(0,d): \phi(0)=0, \ \ \phi(d)=\theta \}, \ \ n=0,1,2,
\end{split}
\]
and $H^n(0,d)$ is the Sobolev space of functions on the interval $(0,d)$. The norms in these spaces will be denoted by $\|\cdot,H^n\|$ for $n=1,2$ and $\|\cdot,L^2\|$ for $n=0$.

Using the identity $\eta_x = -\hat{\Psi}(x,d)/\hat{\Phi}_z(x,d)$ which follows from the definition of $\hat{\Psi}$ by letting $z = d$, the Bernoulli's equation \eqref{bep1} becomes
\begin{equation} \label{eqBP}
\hat{\Psi}^2(x,d) + \hat{\Phi}_z^2(x,d) = P(\eta(x)),
\end{equation}
where the function $P(t)$ is defined by
\[
P(t) = t^2[2Q-2t]/d^2.
\]
Note that the problem \eqref{tmp1}-\eqref{eqBP} depends on $\lambda$, because $\omega, m, Q$ and $u$ do. \\

\subsection{Linearization}

In order to linearize equations near the stream solution $(u(z;\lambda),d(\lambda))$, we put
\begin{equation}\label{K10a}
\hat{\Phi} = \bar{\Phi} + u, \ \ \hat{\Psi} = \bar{\Psi}, \ \ \eta = \zeta + d.
\end{equation}
Thus, equations \eqref{tmp1} and \eqref{tmp2} together with the boundary conditions \eqref{boc1} and \eqref{eqBP} lead to
\begin{align}
& \bar{\Phi}_x  = \bar{\Psi} + \frac{zu_z \zeta_x}{d} + \bar{N}_1 \label{eq:Phi}\\
& \bar{\Psi}_x  = -\bar{\Phi}_{zz} - \frac{2\zeta \omega(u)}{d} - \omega'(u) \bar{\Phi} + \bar{N}_2 \label{eq:Psi}  \\
& \bar{\Phi}(x,0) = \bar{\Phi}(x,d) = \bar{\Psi}(x,0) = 0 \label{eq:PhiB1}\\
& \bar{\Phi}_z(x,d) - \left( \frac{k}{d} - \frac{1}{k} \right) \zeta = \bar{N}_{3}, \label{eq:PhiB}
\end{align}
where the nonlinear operators $\bar{N}_j,j=1,2$, are given by
\[
\begin{split}
  \bar{N}_1(\bar{\Psi}, \bar{\Phi}_z; \zeta, \zeta_x) =  -\frac {\zeta \bar{\Psi}} {d + \zeta}  + z \zeta_x \left[\frac{d \bar{\Phi}_z - \zeta u_z}{d(d+\zeta)} \right] \\
 \end{split}
\]
and
\begin{equation}
\begin{split}
\ \ \ \bar{N}_2(\bar{\Psi}, & \bar{\Psi}_z,  \bar{\Phi},\bar{\Phi}_{zz}; \zeta,  \zeta_x) = \\
& \frac{\zeta_x}{d+\zeta} (z\bar{\Psi})_z + \frac {\zeta (\bar{\Phi}_{zz}-\omega'(u) \bar{\Phi})} {d + \zeta}  - \frac{\zeta^2}{d(d+\zeta)} u_{zz}   \\
&  - \frac{\zeta+d}{d} (\omega(u + \bar{\Phi}) - \omega(u) - \omega'(u) \bar{\Phi}),
\end{split}\label{N2}
\end{equation}
while the nonlinear part in the Bernoulli's equation is defined by
\[
\bar{N}_3(\bar{\Psi}, \bar{\Phi}_z; \zeta) = \frac{1}{2k} \left[ -(\bar{\Psi}^2 + \bar{\Phi}_z^2)_{z=d} + (P(d+\zeta) - P(d) - P'(d)\zeta) \right].
\]
Next we use the following change of variables (see \cite{EhrnstromEscherWahlen11}, Sect.3):
\begin{equation}\label{K10b}
\Phi = \bar{\Phi} - \frac{z u_z \zeta}{d}, \ \ \Psi = \bar{\Psi}.
\end{equation}
Then the relations $\bar{\Phi}(x,d) = 0$ and $\eta_x = -\Psi(x,d)/\hat{\Phi}_z(x,d)$ imply
\begin{equation}\label{zetax}
\zeta = -(1/k) \Phi(x,d), \ \ \zeta_x = - \frac{\Psi(x,d)}{k + \Phi_z(x,d) - \left[ \frac{1}{d} - \frac{\omega(1)}{k} \right] \Phi(x,d)}.
\end{equation}
for all $x \in \RR$. This allows to rewrite equations \eqref{eq:Phi}-\eqref{eq:PhiB} as a first order system in terms of the functions $\Psi$ and $\Phi$ only:
\begin{align}
& \Phi_x  = \Psi + \hat{N}_1 \label{eq:Phihat}\\
& \Psi_x  = -\Phi_{zz} - \omega'(u) \Phi + \hat{N}_2 \label{eq:Psihat}\\
& \Phi(x,0) = \Psi(x,0) = 0 \label{eq:PhiB1hat} \\
& \Phi_z(x,d) - \kappa \Phi(x,d) = \hat{N}_{3}, \label{eq:PhiBhat}
\end{align}
where $\kappa$ is given by (\ref{kappadef}).
The nonlinear operators $\hat{N}_j$ are naturally defined by
\begin{eqnarray} \label{DefN}
 &&\hat{N}_1(\Phi,\Psi) =    -\frac {\zeta \Psi} {d + \zeta}  + z \zeta_x \frac{d \Phi_z +z \zeta u_{zz}}{d(d+\zeta)} ;\nonumber\\
&& \hat{N}_2(\Phi,\Psi) =\frac{\zeta_x}{d+\zeta} (z\Psi)_z
 + \frac {\zeta (\Phi_{zz}-\omega'(u) \Phi-2\zeta z\omega'(u)u_z/d)} {d + \zeta}   \nonumber   \\
&&  - \frac{\zeta+d}{d} \Big(\omega\Big(u + \Phi+\frac{z u_z \zeta}{d}\Big) - \omega(u) - \omega'(u) \Big(\Phi+\frac{z u_z \zeta}{d}\Big)\Big),\nonumber \\
&& \hat{N}_3(\Phi,\Psi) =\frac{1}{2k} \left[ -\Big(\Psi^2 + \Big(\Phi_z+\frac{(z u_z)_z \zeta}{d}\Big)^2\Big)_{z=d} + (P(d+\zeta) - P(d) - P'(d)\zeta) \right],\nonumber
\end{eqnarray}
where $\zeta = \zeta( \Phi)$ and $\zeta_x = \zeta_x(\Phi, \Psi)$ are defined by \eqref{zetax}.

According to the definition, we have that $\Phi$ and $\Psi$ are continuous functions of $x$-variable with values in the Hilbert spaces $\HH^2_{b}$ and $\HH^1_b$, respectively. In this case $\zeta$ and $\zeta_x$ are analytic function of $\Phi$ and $\Psi$ and are of order $C^{3+\nu,\gamma}$ and $C^{2+\nu,\gamma}$ with respect to $\lambda \in \Lambda$. Thus, we have $\hat{N}_1, \hat{N}_2 \in C^{\nu+1}(\HH^2_{b} \times \HH^1_{b} \times \Lambda; L^2(0,d))$, while $\hat{N}_3 \in C^{\nu+1}(\HH^2_{b} \times \HH^1_{b} \times \Lambda; \RR)$. Furthermore, all the derivatives of the latter operators are bounded and uniformly continuous. Thus, the problem \eqref{eq:Phihat}-\eqref{eq:PhiBhat} can be seen as an evolutionary system, where the role of time is played by $x$-variable. Furthermore, the latter dynamical system is reversible, with the reverser defined by $(\Phi, \Psi) \mapsto (\Phi,-\Psi)$.

Next, by using the implicit function theorem we solve equation  (\ref{eq:PhiBhat}) with respect to ${\Phi}_z$ and obtain
\begin{equation}\label{BC1}
\Phi_z(x,d) - \kappa \Phi(x,d)=F_3(\Psi, \Phi),
\end{equation}
where $F_3$ is an analytic function satisfying
$$
F_3(\xi_1, \xi_2)=O(|\xi|^2).
$$
Moreover, one can verify, for example by solving (\ref{eq:PhiBhat}) with the help of fixed point iterations, that
\[
F_3(\Psi, \Phi)=\hat{N}_3(\Psi,\Phi, \kappa \Phi)+O(|\Psi|^3+|\Phi|^3).
\]

\subsection{Spectral decomposition}

According to our notations $\mu = (\mu_1,...,\mu_N)$ stands for the first $N$ eigenvalues of the Sturm-Liouville problem \eqref{SLdisp} for the triple $(\omega(\cdot,\lambda), d, u(\cdot,\lambda))$. Note that by construction these eigenvalues are non-positive for $\lambda=\lambda^\star$, while all others are positive for $\lambda\in\Lambda$. 
We define projectors
$$
{\cal P}_{\lambda}\phi = \sum_{j=1}^N \alpha_j \varphi_j,\;\;\alpha_j = \int_0^d \phi \varphi_j \,dz, \ \ \ \widetilde{\cal P}_{\lambda} = {\rm id} - {\cal P}.
$$
The projectors ${\cal P}_{\lambda}$ and $\widetilde{\cal P}_{\lambda}$ are orthogonal  in $L^2(0,d)$ and are well defined operators on $H^n(0,d)$, $n=1,2$. Note that the set of eigenfunctions $\{\phi_j\}_{j \in \mathbb{N}}$ is a basis in the spaces $L^2(0,d)$ and $H^1_b$. In general the eigenfunction are orthogonal with respect to the following bilinear form
\[
{\cal B}(\phi,\psi) = \int_0^d [\phi' \psi' - \omega'(u) \phi \psi] dz - \kappa \phi(d) \psi(d),
\]
which is well defined on $H^1_{b} \times H^1_{b}$. Moreover, one can verify that
\[
{\cal B}(\phi,\phi_j) = \mu_j \langle \phi, \phi_j \rangle, \ \ j \in \mathbb{N}, \ \ \phi \in H^{1}_0.
\]
We represent the functions $\Psi$ and $\Phi$ as
\begin{equation}\label{KKK11a}
\begin{split}
& \Phi={\cal P}_{\lambda}\Phi+\widetilde{\cal P}_{\lambda}\Phi=\sum_{j=1}^N\alpha_j(x)\varphi_j+\widetilde{\phi}, \\
& \Psi={\cal P}_{\lambda}\Psi+\widetilde{\cal P}_{\lambda}\Psi=\sum_{j=1}^N\beta_j(x)\varphi_j+\widetilde{\psi}.
\end{split}
\end{equation}
By the definition the functions $\widetilde{\Phi}(x,\cdot),\widetilde{\Psi}(x,\cdot)$ are orthogonal in $L^2(0,d)$ to the functions $\varphi_j$ for all $x \in \RR$ and all $j=1,...,N$.

Multiplying \eqref{eq:Phihat}-\eqref{eq:Psihat} by $\varphi_j$, $j=1,\ldots,N$, and integrating over $(0,d)$, we get
\begin{align}
 & \alpha_j' = \beta_j + F_{1j} \label{vtmp1} \\
 & \beta_j' = \mu_j \alpha_j +  F_{2j} \label{vtmp2};
\end{align}
where
$$
F_{1j}=\int_0^d\hat{N}_1(\Psi,\Phi,\Phi_z)\varphi_j dz,\;\;
$$
and
$$
F_{2j}=\int_0^d\hat{N}_2(\Psi,\Phi,\Phi_z,\Phi_{zz})\varphi_j dz-F_3(\Psi,\Phi)\varphi_j(d),\;\;
$$
Subtracting the sum of equations (\ref{vtmp1}) and (\ref{vtmp2}) multiplied by $\varphi_j$ from
(\ref{eq:Phihat}) and (\ref{eq:Psihat}) respectively, we obtain
\begin{align}
 \ \ \ \ \ \ \ \ \ \ \ \ \ \ \ \ & \widetilde{\phi}_x = \widetilde{\psi} + \widetilde{\cal P}_{\lambda}(\hat{N}_1) \label{vtmp3} \\
& \widetilde{\psi}_x = - \widetilde{\phi}_{zz} - \omega'(u) \widetilde{\phi} + \widetilde{\cal P}_{\lambda}(\hat{N}_2)+\sum_{j=1}^NF_3(\Psi,\Phi)(d)\varphi_j(d)\varphi_j. \nonumber \label{vtmp4}
\end{align}
The boundary conditions (\ref{eq:PhiB1hat}) and (\ref{BC1}) take the form
\begin{equation}\label{vtmp4a}
\widetilde{\phi}(x,0)=\widetilde{\psi}(x,0)=0,\;\;\;\widetilde{\phi}_z(x,d) - \kappa \widetilde{\phi}(x,d) = F_{3}(\Psi,\Phi).
\end{equation}
By definition, we have $\widetilde{\phi} \in \widetilde{{\cal H}}_\lambda^2$ and $\widetilde{\psi} \in \widetilde{{\cal H}}_\lambda^1$, where
\[
\widetilde{{\cal H}}_\lambda^n=\{\widetilde{\psi}\in H^n_0 \,:\, {\cal P}_\lambda\widetilde{\psi}=0\},\;\;n=1,2.
\]
It will be also convenient to put
\[
\widetilde{{\cal H}}_\lambda^0 = \{\widetilde{\psi}\in L^2(0,d)\,:\, {\cal P}_\lambda\widetilde{\psi}=0\}
\]
Now we can formulate our central theorem.
\subsection{Main result}

\begin{theorem}\label{T1} There exist neighborhoods $W$, $W_2$ and $W_1$ of the origins in $\Bbb R^{2N}$, $\widetilde{{\cal H}}^2_{\lambda}$ and in $\widetilde{{\cal H}}^1_{\lambda}$ respectively and a smooth vector functions
$$
h\;:\; W\times\Lambda\to W_2,\;\;\;g\;:\; W\times\Lambda\to W_1
$$
with the following properties: \\

1). The functions $h$ and $g$ are of the class $C^{\nu}(W \times \Lambda)$ with values in $W_2$ and $W_1$ respectively, while the corresponding derivatives are bounded and uniformly continuous. Moreover,
$$
||h;H^2||+||g;H^1||=O(|\alpha|^2+|\beta|^2).
$$

2). We introduce the system
\begin{eqnarray}\label{KK9a}
 && \alpha_j' = \beta_j + f_{1j}(\alpha,\beta) \nonumber \\
 && \beta_j' = \mu_j^2 \alpha_j +  f_{2j}(\alpha,\beta),\;j=1,\ldots,N.
\end{eqnarray}
Here
$$
f_{1j}(\alpha,\beta;\lambda)=F_{1j}(\Psi,\Phi,\Phi_z)\,\;\mbox{and}\;\;
f_{2j}(\alpha,\beta;\lambda)=F_{2j}(\Psi,\Phi,\Phi_z,\Phi_{zz}),
$$
where $F_{1j}$ and $F_{2j}$ are the same as in system (\ref{vtmp1}), (\ref{vtmp2}) and
\[
\begin{split}
& \Phi(\alpha,\beta; \lambda)(x,z)=\sum_{j=1}^N \alpha_j(x)\varphi_j(z)+h(\alpha,\beta;\lambda)(x,z), \\
& \Psi(\alpha,\beta; \lambda)(x,z)=\sum_{j=1}^N\beta_j(x)\varphi_j(z)+g(\alpha,\beta;\lambda)(x,z).
\end{split}
\]
Let us also define
\[
M^\lambda=\{ \hat{\Phi}(\alpha,\beta; \lambda),\hat{\Psi}(\alpha,\beta; \lambda))\,:\,(\alpha,\beta)\in W, \ \lambda \in \Lambda \},
\]
where
\[
\hat{\Phi}(x,z)=u(z)-\frac{zu_z(z)\Phi(x,d)}{kd}+\Phi(x,z),\;\; \hat{\Psi}(x,z) = \Psi(x,z)
\]
Then
(i) $M^\lambda\subset H^2_{0,m} \times H^1_{0}$ is a locally invariant manifold of (\ref{tmp1})-(\ref{tmp2}): through every point in $M^\lambda$ there passes a unique solution of (\ref{tmp1})-(\ref{tmp2}) that remains on $M^\lambda$ as long as $(h,g)$ remains in $W_2\times W_1$;
(ii) every  bounded solution $(\hat{\Phi}, \hat{\Psi})$ of (\ref{tmp1})-(\ref{eqBP}) for which $(\alpha,\beta)\in W$ and $(h,g)\in W_2\times W_1$ lies completely in $M^\lambda$, provided the norm $\|\hat{\Phi}-u; H^2\|+\|\hat{\Psi};H^1\|$ is small;
(iii) every solution $(\alpha,\beta)\,:\,(a,b)\rightarrow W$ of the reduced system (\ref{KK9a}) generates a solution $(\hat{\Phi}, \hat{\Psi}, \eta)$ of the full problem (\ref{tmp1})-(\ref{eqBP}), where $\Phi$ and $\Psi$ are defined by as above and $\eta = d - [\Phi]_{z=d}/k$; (iv) the reduced system (\ref{KK9a}) is reversible. \\
\end{theorem}

The proof of this theorem is given in the next three sections.

\subsection{Change of variables}

The proof of the theorem is based on the application of a reduction theorem due to Mielke. However, we can not apply Mielke's result directly to the system \eqref{vtmp1}-\eqref{vtmp4a} because of the nonlinear boundary condition \eqref{vtmp4a}. We overcome this difficulty by passing to a new variables for which all  boundary conditions are homogeneous.

First, we put ${\cal P}_*={\cal P}_{\lambda_*}$ and
$\widetilde{\cal P}_*=\widetilde{\cal P}_{\lambda_*}$. By a new change of variable we will reduce the last boundary condition in (\ref{BC1}) to a homogeneous one (without nonlinear term), which is independent of the parameter $\lambda$. The change of variable
 $$
 \widetilde{{\cal H}}^2_\lambda\times \widetilde{{\cal H}}^1_\lambda\ni (\widetilde{\phi},\widetilde{\psi})\rightarrow (w,v)\in  \widetilde{{\cal H}}^2_{\lambda_\star}\times \widetilde{{\cal H}}^1_{\lambda_\star}
 $$
  is the following (compare with \cite{BuffoniGrovesToland96}, \cite{GrovesWahlen08}):
\begin{eqnarray}
&&v=\widetilde{\cal P}_*\widetilde{\psi},\;\;\; w=\widetilde{\cal P}_*\widetilde{\Phi}\label{KK8}\\
&&\widetilde{\Phi}=\widetilde{\phi}+\widetilde{\cal P}_* \frac{z}{d}\int_z^{d}\Big(F_3(\Psi,\Phi)(\tau)+(\kappa-\kappa_*)\widetilde{\phi}(\tau)\Big)d\tau.\label{KK8w}
\end{eqnarray}
Here
\begin{equation*}
\Phi=\phi+\widetilde{\phi},\;\;\Psi=\psi+\widetilde{\psi},\;\;\mbox{and}\;\;
\phi=\sum_{j=1}^N \alpha_j \varphi_j,\;\;\psi=\sum_{j=1}^N\beta_j\varphi_j.
\end{equation*}
One can verify directly that the function $\widetilde{\Phi}$, and hence $w$ satisfies the boundary condition
\begin{equation}\label{KK13c}
[\widetilde{\Phi}_y-\kappa_*\widetilde{\Phi}]_{z=d}=0
\end{equation}
 if and only if $\widetilde{\phi}$ satisfies the last boundary condition in (\ref{vtmp4a}). Our aim is to invert relations (\ref{KK8}) and (\ref{KK8w}) and to express $\widetilde{\psi}$ and $\widetilde{\phi}$
 through  $v$, $w$ and $\alpha$, $\beta$, $\lambda$. This can be done directly by means of the implicit function theorem, however we are also interested in explicit formulas giving the leading terms of the approximation. First, we find  the inverse to the operator
\[
\widetilde{{\cal H}}_\lambda^n\ni\widetilde{\psi}\rightarrow \widetilde{\cal P}_*\widetilde{\psi}=v\in \widetilde{\cal H}_{\lambda_*}^n.
\]
We are looking for it in the form
\begin{equation}\label{KK8a}
\widetilde{\psi}=\widetilde{\cal P}_\lambda(I+S_\lambda)v,
\end{equation}
where $S_\lambda\,:\,\widetilde{\cal H}_{\lambda_*}^n\rightarrow \widetilde{\cal H}_{\lambda_*}^n$. Then substituting \eqref{KK8a} in \eqref{KK8}, we find
\[
v=\widetilde{\cal P}_*\widetilde{\psi}=(I-\widetilde{\cal P}_*({\cal P}_\lambda-{\cal P}_*))(I+S_\lambda)v.
\]
Therefore, we express
\[
S_\lambda v=\widetilde{\cal P}_*({\cal P}_\lambda-{\cal P}_*)(I+S_\lambda)v.
\]
Note that the operator $I-\widetilde{\cal P}_*({\cal P}_\lambda-{\cal P}_*)$ is invertible, provided  $\lambda$ is close to $\lambda_*$. Thus, we can resolve $S_\lambda v$ from the last identity, which gives
\[
S_\lambda v=\big(I-\widetilde{\cal P}_*({\cal P}_\lambda-{\cal P}_*)\big)^{-1}\widetilde{\cal P}_*({\cal P}_\lambda-{\cal P}_*)v.
\]
This allows us to resolve the first relations in (\ref{KK8}):
\begin{equation}\label{KK8b}
\widetilde{\psi}=\widetilde{\cal P}_\lambda\Big(I+\big(I-\widetilde{\cal P}_*({\cal P}_\lambda-{\cal P}_*)\big)^{-1}\widetilde{\cal P}_*({\cal P}_\lambda-{\cal P}_*)\Big)v.
\end{equation}
 The operator $S_\lambda\, :\,\widetilde{\cal H}_{\lambda^*}^n\rightarrow \widetilde{\cal H}_{\lambda^*}^n$ is continuous with the norm of the order $O(|\lambda-\lambda_*|)$.
Solving the second equation in (\ref{KK8}), we have
\begin{equation}\label{KK8bb}
\widetilde{\Phi}=\widetilde{\cal P}_\lambda\Big(I+\big(I-\widetilde{\cal P}_*({\cal P}_\lambda-{\cal P}_*)\big)^{-1}\widetilde{\cal P}_*({\cal P}_\lambda-{\cal P}_*)\Big)w.
\end{equation}
We can also write relations (\ref{KK8b}) and (\ref{KK8bb}) as
\[
\widetilde{\psi}=v+M(\lambda)v\;\;\mbox{and}\;\;\widetilde{\Phi}=w+M(\lambda)w,
\]
where
$$
M(\lambda)=\Big (I-\widetilde{\cal P}_\lambda\big(I-\widetilde{\cal P}_*({\cal P}_\lambda-{\cal P}_*)\big)^{-1}\widetilde{\cal P}_*\Big)({\cal P}_*-{\cal P}_\lambda).
$$
The operator function $M$ is $(\nu+1)$ times differentiable and of order $O(|\lambda-\lambda_*|^2)$.

Let us solve  equation (\ref{KK8w}) with respect to $\widetilde{\phi}$. 
We are looking for the solution in the form
$$
\widetilde{\phi}=\widetilde{\Phi}+R,\;\;\;R=R(\widetilde{\Phi};v,\alpha,\beta,\lambda),
$$
where $R$ is a nonlinear operator defined in a neighborhood of the origin in ${\cal H}^2_{\lambda_*}$. Substituting this in to  (\ref{KK8w}),
we get
\begin{equation}\label{KK8c}
R=-\widetilde{\cal P}_* \frac{z}{d} \int_z^{d}\Big(F_3(\Psi,\Phi)(\tau)+(\kappa -\kappa_*)(\widetilde{\Phi}(\tau)+R)\Big)d\tau.
\end{equation}
Here one must put
$$
\Psi=\psi+{\cal P}_\lambda(I+S)v,\;\;\Phi=\phi+\widetilde{\Phi}+R.
$$
Applying  the fixed point theorem to equation (\ref{KK8c}) we find $R$ as the function of $(\widetilde{\Phi},v,\alpha,\beta,\lambda)$. Moreover, if the function
$F_3$ is $C^{\nu+1}$ smooth with respect to these arguments the same is true for the function $R$. Furthermore, if $\widetilde{\phi}$ satisfies the relation (\ref{BC1}) then the function $\widetilde{\Phi}$ satisfies (\ref{KK13c}).


Thus, we obtain
\begin{equation}\label{K11a}
\widetilde{\psi}=v+M(\lambda)v\;\;\mbox{and}\;\;\widetilde{\phi}=w+M(\lambda)w+R(v,w;\alpha,\beta,\lambda),
\end{equation}
where  $R$ is $C^{\nu+1}$ function in a neighborhood of $(0,0,0,0,\lambda_*)$ in the space $\widetilde{\cal H}^1_{\lambda_*}\times \widetilde{\cal H}^2_{\lambda_*}\times\Bbb R^N\times\Bbb R^N\times\Lambda$ with values in $\widetilde{\cal H}^2_{\lambda_*}$.
We can represent also $R$ as $R_1+Q$, where $Q$ satisfies
\[
Q=-\widetilde{\cal P}_* \frac{z}{d}\int_z^{d}(\kappa-\kappa_*)(\widetilde{\Phi}(\tau)+Q)d\tau
\]
and $R_1$ solves the equation
\[
R_1=-\widetilde{\cal P}_* \frac{z}{d}\int_y^{d}\Big(F_3(\Psi,\Phi)(\tau)+(\kappa-\kappa_*)R_1\Big)d\tau.
\]
Then $Q=Q(w;\lambda)$ is a linear operator with respect to $w$ satisfying
$$
||Q(w;\lambda); H^2||\leq C|\lambda-\lambda_*|\;||w;H^1||
$$
and
 $R_1(v,w;\alpha,\beta,\lambda)$ is $C^{\nu+1}$  function mapping a neighborhood of $(0,0,0,0,\lambda_*)$ to
 $\widetilde{\cal H}^2_{\lambda_\star}$ and
\begin{equation*} ||R_1(v,w;\alpha,\beta,\lambda);H^2||=O\Big(||v;H^1||^2+||w;H^2||^2+|\alpha|^2+|\beta|^2\Big).
\end{equation*}
The representation (\ref{K11a}) takes the form
\[
\widetilde{\psi}=v+M(\lambda)v\;\;\mbox{and}\;\;\widetilde{\phi}=w+M(\lambda)w+Q(w;\lambda)+R_1(v,w;\alpha,\beta,\lambda).
\]
After this change of variables the problem (\ref{vtmp3})-(\ref{vtmp4a}) becomes
\begin{eqnarray}\label{KK14a}
&& w_x = v + \widetilde{\cal P}_*f_1(w,v,\alpha,\beta,\lambda)  \nonumber\\
&& v_x = - w_{yy} - \omega_\star'(u) w + \widetilde{\cal P}_*(Q_1(w;\lambda)+f_2(w,v,\alpha,\beta,\lambda))
\end{eqnarray}
and
\begin{equation}\label{KK14b}
w(x,0)=v(x,0)=0,\;\;\;w_y(x,d_*) - \kappa_\star w(x,d_*) = 0.
\end{equation}
Here $Q_1(w;\lambda)$ is a linear operator with respect to $w$ satisfying
$$
||Q_1(w;\lambda); H^2||\leq C|\lambda-\lambda_*|\;||w;H^2_{0}||.
$$
To describe required properties of $f_1$ and $f_2$ let
$$
B^1_{\varepsilon}=\{v\in \widetilde{\cal H}^1_{\lambda_\star}\,:\, ||v;H^1||\leq \varepsilon\},\;\;B^2_{\varepsilon}=\{w\in \widetilde{\cal H}^2_{\lambda_\star}\,:\, ||v;H^2||\leq \varepsilon\}
$$
and
$$
B_\varepsilon(a;M)=\{X\in\Bbb R^{M}\,:\,|X-a|\leq\varepsilon\}.
$$
Then for a sufficiently small $\varepsilon > 0$, we have
$$
f_1\,:\,B^2_\varepsilon\times B^1_\varepsilon\times B_\varepsilon (0;2N)\times B_\varepsilon (\lambda_\star;m)\to H^1_{0},
$$
$$
f_2, \;Q\,:\,B^2_\varepsilon\times B^1_\varepsilon\times B_\varepsilon (0;2N)\times B_\varepsilon (\lambda_\star;m)\to L^2(0,d)
$$
are $C^{\nu+1}$ times continuously differentiable  maps, satisfying
\begin{equation*}
||f_1;H^1||+||f_2;L^2||=O(||v;H^2||^2+||w;H^1||^2+|\alpha|^2+|\beta|^2).
\end{equation*}

\subsection{Mielke's reduction theorem}

Let ${\cal X}$ be a Hilbert space which is represented as a product ${\cal X}={\cal X}_1\times {\cal X}_2$ of two Hilbert spaces ${\cal X}_1$ (finite dimensional) and ${\cal X}_2$ (infinite dimensional) with the norms $||\cdot||$ and $||\cdot||_2$ respectively.
Consider the following  system of differential equations
\begin{equation}\label{KK4a}
\dot{x}_1=A(\lambda)x_1+f_1(x_1,x_2,\lambda),
\end{equation}
\begin{equation}\label{KK4b}
\dot{x}_2=Bx_2+Q(x_1,x_2,\lambda)x_2+f_2(x_1,x_2,\lambda),
\end{equation}
where $A(\lambda)$ is a linear operator in ${\cal X}_1$ and $B\,:\, {\cal D}\subset {\cal X}_2\rightarrow{\cal X}_2$ is a  closed linear operator. We will consider ${\cal D}$ as a Hilbert space supplied with the graph norm $||x_2;{\cal D}||=(||x_2||_2^2+||Bx_2||_2^2)^{1/2}$. Here $\lambda$ is a parameter located in a neighborhood $\Lambda'\subset\Bbb R^m$ of a point $\lambda_0$;
\begin{eqnarray*}\label{KK4ab}
&&f_1\,:\;{\cal U}_1'\times {\cal U}_2'\times\Lambda'\to{\cal X}_1,\;\;\;
f_2\,:\;{\cal U}_1'\times {\cal U}_2'\times\Lambda'\to{\cal X}_2\;\;\;\mbox{and}\nonumber\\
&&Q\,:\;{\cal U}_1'\times {\cal U}_2'\times\Lambda'\to{\cal X}_2
\end{eqnarray*}
 are continuously differentiable functions. Here ${\cal U}_1'$ and ${\cal U}_2'$ are neighborhoods of the origin in the spaces ${\cal X}_1$ and ${\cal D}$ respectively.

We assume that

(A1) the operator function $A(\lambda)$ is $(\nu+1)$ times continuously differentiable with respect to $\lambda$ and the spectrum of $A(\lambda_0)$ lies on the imaginary axis;

(A2) the operator $B\,:\,{\cal D}\rightarrow{\cal X}_2$ is continuous operator and for all $\xi\in\Bbb R$ the operator $B-i\xi$ is invertible and
\[
||(B-i\xi)^{-1}||\leq \frac{C}{1+|\xi|}
\]
for some constant $C$ independent of $\xi$;

(A3) the functions $f_1,f_2,Q$ are $(\nu+1)$ times continuously differentiable and their derivatives are bounded and uniformly continuous on the corresponding definition domains. Moreover,
\begin{eqnarray}\label{KK4d}
&&||f_1(x_1,x_2,\lambda)||=O(||x_1||^2+||x_2;D||^2), \nonumber\\
&&||f_2(x_1,x_2,\lambda)||_2=O(||x_1||^2+||x_2; D||^2),\nonumber\\
&&||Q(x_1,x_2,\lambda)||_2=O\big(||x_1||+||x_2;D||+|\lambda-\lambda_0|\big)
\end{eqnarray}
for $(x_1,x_2,\lambda)\in {\cal U}'_1\times {\cal U}'_2\times\Lambda$.

Then there exist neighborhoods
$$
{\cal U}_1\subset {\cal U}_1',\;\;{\cal D}_2\subset {\cal D}_2',\;\;\Lambda\subset\Lambda'
$$
of $0$, $0$ and $\lambda_0$ respectively and a reduction function
\begin{equation}\label{KK4e}
h\,:\,{\cal U}_1\times\Lambda\rightarrow D
\end{equation}
 with the following properties: the function (\ref{KK4e}) is $\nu$ times continuously differentiable and its derivatives are bounded and uniformly continuous on
 ${\cal U}_1\times\Lambda$ and
\begin{equation}\label{KK4f}
h(x_1,\lambda)=O(||x_1||^2)\;\;\;\mbox{for all $\lambda\in\Lambda$}.
\end{equation}
The graph
$$
M_C^\lambda=\{(x_1,h(x_1,\lambda)\in {\cal U}_1\times {\cal D}_2\,:\, x_1\in {\cal U}_1\}
$$
is a center manifold for (\ref{KK4a}), (\ref{KK4b}), which means that:

(1) every small bounded solution of (\ref{KK4a}), (\ref{KK4b}) with $x_1(t)\in {\cal U}_1$ and $x_2(t)\in D_2$ lies completely in $M_C^\lambda$;

(2) every solution $x_1(t)$, $t\in\Bbb R$, of the reduced equation
$$
\dot{x}_1=A(\lambda)x_1+f_1(x_1,h(x_1,\lambda),\lambda)
$$
generates a solution
$(x_1(t),x_2(t))$, $x_2(t)=h(x_1(t),\lambda)$ of the equation (\ref{KK4a}), (\ref{KK4b}).

\bigskip
This theorem is taken basically from \cite{Mielke1991}, see also \cite{Mielke88}, \cite{Mielke1986}, \cite{BuffoniGrovesToland96}, \cite{GrovesWahlen08}. The only small difference is that we split the right-hand side in (\ref{KK4a}), (\ref{KK4b}) into linear parts with respect to $x_1$ and $x_2$ and quadratically depending on $x_1$ and $x_2$. This allows us to write more explicit estimate (\ref{KK4f}) for the reduction function $h$. The proof of this improvement is quite straightforward and we omit it.

\subsection{Proof of Theorem \protect\ref{T1}}

We will apply Mielke's theorem to the problem (\ref{vtmp1}), (\ref{vtmp2}), (\ref{KK14a}), (\ref{KK14b}).
In order to do this we choose
$$
{\cal X}_1=\Bbb R^{2N},\;\;\;{\cal X}_2=\widetilde{\cal H}^1_{\lambda_*}\times \widetilde{\cal H}^0_{\lambda_*}\;\;\mbox{and}\;\;{\cal D}=\widetilde{\cal H}^2_{\lambda_*}\times \widetilde{\cal H}^1_{\lambda_*}.
$$

The property (A1) follows from (\ref{KK12a}).

 In order to verify the property (A2) in the Mielke's theorem let us consider the problem
\begin{align}
& i\xi w  -v = f \label{KK2a}\\
& i\xi v  + w_{zz} + \omega_\star'(u_\star) w = g, \label{KK2b}
\end{align}
where $(f,g)\in {\cal X}_2$ and $(v,w)\in D$.
 We are looking for solution to problem (\ref{KK2a}), (\ref{KK2b}) in the form
$$
w=\sum_{j=N+1}^\infty a_j\phi_j,\;\;\;v=\sum_{j=N+1}^\infty b_j\phi_j.
$$
We represent also
$$
f=\sum_{j=N+1}^\infty f_j\phi_j,\;\;\;g=\sum_{j=N+1}^\infty g_j\phi_j.
$$
Then
$$
b_j=-\frac{\mu_j}{\mu_j+\xi^2}\Big(f_j+\frac{i\xi}{\mu_j}g_j\Big),\;\;a_j=\frac{-i\xi}{\mu_j+\xi^2}f_j-\frac{1}{\mu_j+\xi^2}g_j
$$
Since the norms $||w;H^1||$ and $||v;L^2||$ are equivalent to the norms
$$
\Big(\sum_{j=N+1}^\infty \mu_ja_j^2\Big)^{1/2}\;\;\mbox{and}\;\;\Big(\sum_{j=N+1}^\infty b_j^2\Big)^{1/2}
$$
respectively, we verify that
\[
||w;H^1_{b}||^2+||v;L^2||^2\leq\frac{C}{1+\xi^2}\Big(||f;H^1_{b}||^2+||g;L^2||^2\Big).
\]

\smallskip
(A3) We require that the vorticity function $\omega(y;\lambda)$ is of class $C^{\nu+2}$, which implies $(\nu+1)$ times continuous differentiability of the functions $\hat{N}_1$, $\hat{N}_2$ and $\hat{N}_3$.
The relations (\ref{KK4d}) are true for the functions $N_j$, $j=1,2,3$, and they preserve after changes of variables.

\smallskip
Now the application of Mielke's theorem gives the existence of the reduction vector function
$$
(h,g)\;:\;W\times\Lambda\rightarrow W_2\times W_1,
$$
which reduces the problem (\ref{vtmp1}), (\ref{vtmp2}), (\ref{KK14a}), (\ref{KK14b}) to the fine dimensional system
(\ref{KK9a}). Now returning to system (\ref{tmp1}), (\ref{tmp2}) with boundary conditions (\ref{boc1}) with the help of transformations (\ref{KK8}), (\ref{KK8w}), (\ref{KKK11a}), (\ref{K10b}) and (\ref{K10a}) we complete
the proof of our theorem.

\subsection{Hamiltonian structure of the first order system}
Let us show that the initial problem and the reduced system admit a Hamiltonian structure. For each pair $(\hat{\Phi},\hat{\Psi})\in H^2_{0,m}  \times H^1_{b}$ we define a functional
\begin{equation} \label{symp}
\begin{split}
& \hat{\varpi}_{(\hat{\Phi},\hat{\Psi})} (\hat{\Phi}^{(1)},\hat{\Psi}^{(1)};\hat{\Phi}^{(2)},\hat{\Psi}^{(2)})=\int_0^d(\hat{\Psi}^{(2)}\hat{\Phi}^{(1)}-\hat{\Phi}^{(2)}\hat{\Psi}^{(1)})dz \\
& +\frac{d^2}{\eta^2(2Q-3\eta)}\Big\{[\hat{\Psi}\hat{\Psi}^{(2)}+\hat{\Phi}\hat{\Phi}^{(2)}_{z}]_{z=d}
\int_0^dz(\hat{\Phi}^{(1)}_{z}\hat{\Psi}+\hat{\Phi}_z\hat{\Psi}^{(1)})dz \\
& -[\hat{\Psi}\hat{\Psi}^{(1)}+\hat{\Phi}\Phi^{(1)}_{z}]_{z=d}
\int_0^dz(\hat{\Phi}^{(2)}_{z}\hat{\Psi}+\hat{\Phi}_z\hat{\Psi}^{(2)})dz,
\end{split}
\end{equation}
where $(\hat{\Phi}^{(s)},\hat{\Psi}^{(s)})\in H^2_{0,0}  \times H^1_{b}, \ s=1,2$.
Let us recall the definition of the flow force invariant:
\[
\hat{S}(\hat{\Phi},\hat{\Psi})=\Big [Q+\Omega(1)\Big]\eta-\frac{\eta^2}{2}-\int_0^d\Big[\frac{d}{2\eta}(\hat{\Psi}^2-\hat{\Phi}_z^2)+\frac{\eta}{d}\Omega(\hat{\Phi})\Big]dz.
\]
It was proved in \cite{Kozlov2013} that 
\[
\varpi_{(\hat{\Phi},\hat{\Psi})} (\delta\hat{\Phi},\delta\hat{\Psi};\hat{\Phi}_x,\hat{\Psi}_x)=d\hat{S}(\delta\hat{\Phi},\delta\hat{\Psi})
\]
is a Hamiltonian form of the system (\ref{tmp1})-(\ref{tmp2}) with boundary conditions (\ref{boc1}) and \eqref{eqBP}, where the function $\eta$ is found from (\ref{eqBP}) and $\eta_x=-\hat{\Psi}(x,d)/\hat{\Phi}_z(x,d)$. The disadvantage of \eqref{symp} is that it is well-defined only in a neighbourhood of an equilibrium point (stream solution), provided $\frac{2Q}{3} \neq d$. These assumptions may be omitted if one uses the following change of variables:
\begin{equation}\label{K9a}
\hat{\Phi}=u(z)+\Phi + \frac{zu_z}{d} (\eta - d), \ \ \ \hat{\Psi}=\Psi.
\end{equation}
Thus, using the boundary relation \eqref{boc1}, we can express
\begin{equation} \label{eta}
\eta = d -\frac{[\Phi]_{z=d}}{k}
\end{equation}
and eliminate the profile function from the equations. Now the Bernoulli equation \eqref{eqBP} transforms into
\begin{equation} \label{Bern}
\left[\Psi^2 + \left(u_z + \Phi_z - \frac{(zu_z)_z}{kd}\Phi \right)^2 \right]_{z=d} = P(d - [\Phi]_{z=d}/k).
\end{equation}
Let us define a manifold ${\cal M}$ to be the set of all pairs $(\Phi, \Psi) \in H^2_{b} \times H^1_{b}$ satisfying \eqref{Bern}. Then the tangent space $\rm T{\cal M}_{\Phi,\Psi}$ consists of all pairs $(\Phi^{(s)}, \Psi^{(s)}) \in H^2_{b} \times H^1_{b}$ that are subject to the linear relation
\[
\begin{split}
& \left[ 2 \Psi \Psi^{(s)} +  2 \left(u_z + \Phi_z - \frac{(zu_z)_z}{kd}\Phi \right) \left(\Phi_z^{(s)} - \frac{(zu_z)_z}{kd}\Phi^{(s)} \right) \right]_{z=d} = \\
& -P'(d - [\Phi]_{z=d}/k) \frac{[\Phi^{(s)}]_{z=d}}{k}.
\end{split}
\]
According to the definition \eqref{K9a} and \eqref{eta} we define $(\hat{\Phi}^{(s)},\hat{\Psi}^{(s)}) \in H^2_{0,0} \times H^1_{b}$ by
\[
\hat{\Phi}^{(s)} = \Phi^{(s)} - \frac{zu_z}{kd}[\Phi^{(s)}]_{z=d}, \ \ \hat{\Psi}^{(s)} = \Psi^{(s)}, \ \ s= 1,2;
\]
Plugging it into \eqref{symp} and using \eqref{eta}, we define
\[
 \varpi_{(\Phi,\Psi)} (\Phi^{(1)},\Psi^{(1)};\Phi^{(2)},\Psi^{(2)}) = \hat{\varpi}_{(\hat{\Phi},\hat{\Psi})}(\hat{\Phi}^{(1)},\hat{\Psi}^{(1)};\hat{\Phi}^{(2)},\hat{\Psi}^{(2)}).
\]
Note that the latter definition makes sense without smallness assumption and does not require $\frac{2Q}{3} \neq d$. Now one can show that the system \eqref{eq:Phihat}-\eqref{eq:PhiBhat} can be written as
\[
\varpi_{(\Phi,\Psi)}(\delta\Phi,\delta\Psi;\Phi_x,\Psi_x)=dS(\delta\Phi,\delta\Psi),
\]
where $S(\Phi, \Psi) = \hat{S}(\hat{\Phi},\hat{\Psi})$. Then it follows from Theorem \ref{T1} that this infinite-dimensional Hamiltonian system can be reduced (locally) to a finite-dimensional one corresponding to \eqref{KK9a}. The reduced system is obtained in the following way. We put
\begin{equation} \label{redgraph}
\Phi = \phi + h(\alpha, \beta), \ \ \Psi = \psi + g(\alpha,\beta),
\end{equation}
where
\[
\phi = \sum_{j=1}^N \alpha_j \varphi_j, \ \ \psi = \sum_{j=1}^N \beta_j \varphi_j.
\]
This allows us to write tangent vectors $(\Phi^{(s)},\Psi^{(s)})$ in the form
\begin{equation} \label{deltaPhi}
\begin{split}
\Phi^{(s)} & =\sum_{j=1}^N \alpha_j^{(s)}\varphi_j(z) + \frac{\partial h}{ \partial \alpha} \cdot \alpha^{(s)} + \frac{\partial h}{ \partial \beta} \cdot \beta^{(s)}, \\
\Psi^{(s)} & =\sum_{j=1}^N \beta_j^{(s)}\varphi_j(z)+ \frac{\partial g}{ \partial \alpha} \cdot \alpha^{(s)} + \frac{\partial g}{ \partial \beta} \cdot \beta^{(s)}. \ \
\end{split}
\end{equation}
Thus, the reduced symplectic form is defined by
\[
\vartheta(\alpha^{(1)},\beta^{(1)};\alpha^{(2)},\beta^{(2)}) = \varpi_{(\Phi,\Psi)} (\Phi^{(1)},\Psi^{(1)};\Phi^{(2)},\Psi^{(2)}).
\]
If we introduce the reduced force flow invariand by
$$
s(\alpha,\beta)=S(\hat{\Phi},\hat{\Psi}),
$$
where $\hat{\Phi}$ and $\hat{\Psi}$ are given by \eqref{K9a} with $\Phi$ and $\Psi$ given by \eqref{redgraph}, then
$$
\vartheta(\hat{\alpha},\hat{\beta};\alpha_x,\beta_x)=ds(\hat{\alpha},\hat{\beta}),
$$
which coincides with the reduced system \eqref{KK9a}. It follows directly from the equations that
\begin{equation} \label{s2}
s(\alpha,\beta) = \frac{1}{2} \left[ - \sum_{j=1}^N\beta_j^2 + \sum_{j=1}^N \mu_j \alpha_j^2 \right] + O(|\alpha|^3 + |\beta|^3).
\end{equation}
This implies that the Hamiltonian is a negative definite in a neighborhood of the origin in $\RR^{2N}$, which will be used in the next section.

\section{Non-symmetric steady waves}

One can think of Theorem \ref{T1} as a tool or a framework for the further analysis of the problem. Here we will give a simple proof, in contrast to the known existence results, that most of the solutions provided by Theorem \ref{T1} are not symmetric. First let us explain what we mean by a symmetric solution. Often it means that the surface profile is an even function that is symmetric around $x=0$, but it is natural to identify solutions that can be obtained one from another by a translation in $x$-variable. Thus, we say $(\alpha(x),\beta(x))$ is symmetric if the vector function $(\alpha(x-x_0),\beta(x-x_0))$ is even for some $x_0 \in \RR$.

\begin{theorem} Assume that all the eigenvalues $\mu_j^*$ , $j=1,...,N$ are strictly negative. Then there exists an open neighbourhood $W_*$ of the origin in $\RR^{2N}$ such that the following statements are true:
(i) the reduced system \eqref{KK9a} for $\lambda = \lambda_*$ possesses a unique solution $(\alpha, \beta)$ for any initial data $\alpha(0) = \hat{\alpha}, \beta(0) = \hat{\beta}, \ (\hat{\alpha},\hat{\beta}) \in W_*$; (ii) let $W_*^{sym}$ be the set of all $(\alpha_*,\beta_*) \in W_*$ such that $(\alpha_*,\beta_*) = (\alpha(x),\beta(x))$ for some symmetric solution $(\alpha, \beta)$ and some $x \in \RR$. Then the Hausdorff dimension of $W_*^{sym}$ is less or equal than $N+1$.
\end{theorem}

Thus, if $N \geq 2$, then the most of solutions in $W_*$ are not symmetric. In the case $N=1$ all small-amplitude solutions are symmetric.

\begin{proof} The first part of the statement is trivial since the reduced Hamiltonian is negative definite. Let $W$ be a neighborhood of the origin in $\RR^{2N}$ for which (i) is true. We consider the map
\[
{\mathcal G} : W \times \RR \to \RR^{2N}
\]
defined by
\[
{\mathcal G}(\alpha_*,\beta_*; x) = (\alpha(x),\beta(x)),
\]
where $(\alpha, \beta)$ stands for the unique solution with the initial data $(\alpha_*,\beta_*) \in W$. Because the system is Hamiltonian and in view of \eqref{s2}, we  can choose a neighborhood $\hat{W} \subset W$ such that ${\mathcal G}(\hat{W} \times \RR) \subset W$. Now it follows that
\[
\hat{W}^{sym} \subset {\cal G}(W_0 \times \RR),
\] 
where $W_0 = \{(\alpha,\beta) \in W : \beta = 0\}$. We note that can choose $W$ from the beginning so that ${\mathcal G} \in C^{\nu}(W \times [-T,T])$ for any $T \in \RR$, which implies that ${\mathcal G}$ is lipschitz on every compact subset of $W \times \RR$. Therefore, the Hausdorff dimension of the image ${\mathcal G}(W_0 \times \RR)$ is less or equal than the dimension of $W_0 \times \RR$ which is $N+1$. Hence, we find that $\textrm{dim}_H(\hat{W}^{sym}) \leq N+1$. To finish the proof it is enough to choose $W_* := \hat{W}$.
\end{proof}

\vspace{2mm}

\noindent {\bf Acknowledgements.} V.~K. and E.~L. were supported by the Swedish
Research Council (VR).

\bibliographystyle{siam}
\bibliography{bibliography}

\end{document}